\documentclass[a4paper,11pt]{article}
\pdfoutput=1 % if your are submitting a pdflatex (i.e. if you have
\usepackage{jheppub} % for details on the use of the package, please
\usepackage[T1]{fontenc} 
\usepackage{caption}
\usepackage{subcaption}
\usepackage[english]{babel}
\usepackage{graphicx, color}
\usepackage{hyperref}
\definecolor{link}{rgb}{0.1,0.1,0.9}
\hypersetup{colorlinks=true,linkcolor=link,citecolor=link,urlcolor=link,linktocpage}

\newcommand{\refb}[1]{(\ref{#1})}

\title{\boldmath Two-dimensional Yang-Mills Theory on Recursive Infinite Genus Surfaces}

\author{Dushyant Kumar}

\affiliation{Jawaharlal Nehru University,\\New Delhi 110067, India}
\emailAdd{sehrawat.dushyant@gmail.com}

\abstract{The partition function of Euclidean Yang-Mills theory on two dimensional 
surfaces is given by the Migdal formula. It involves the area and topological 
characteristics of the surface. We consider this theory on a class of infinite genus 
surfaces that are constructed recursively. We make use of this recursive structure 
to compute the partition functions (with or without additional Wilson loops) on such 
surfaces. Our method also works for the quantum deformed Yang-Mills theory.}
\begin{document} 
\maketitle
\flushbottom

%%%%%%%%%%%%%%%%%%%%%%%%%
\section{Introduction}
\label{sec:intro}
Quantum field theories on two-dimensional surfaces (which may have the additional structure of 
Riemann surfaces) are important to understand various aspects of their dynamics\cite{Mig,Wit:first}. 
There are also applications in critical phenomena in two dimensions and in string theory, in which 
conformally invariant theories play a special role. For the latter, an expansion of, say the free energy, 
in terms of the genus of the surface correspond to the quantum loop expansion. The role of surfaces 
of infinite genus have been intriguing from the early days of string theory. Na\"{i}vely one would think 
of the contribution from infinite genus surfaces to correspond to large quantum effects, or
non-perturbative effects. 
However, it is difficult to work with surfaces of infinite characteristics. Therefore,
studies of these have been limited. Among them are the universal moduli space
of Friedan and Shenker~\cite{FS}, and string theory on Hill's surface, a class of hyperelliptic 
Riemann surfaces (HERS)~\cite{MT} studied in Ref.~\cite{LaThermo}. 

In Ref.~\cite{Richards} Richards gave a topological classification of noncompact infinite genus 
surfaces. One of these infinite genus surfaces has its set of `ideal boundaries' (see 
Sec.~\ref{sec:RichCons})  homeomorphic to the Cantor set. As we will describe below, proceeding 
along the lines of Ref. \cite{Richards}, we can construct a family of compactifications of this surface 
that are parametrized by two positive integers. We refer to these compactifications as Richards surfaces.  
The surfaces obtained thus are not smooth. We will make use of the recursive construction of these 
surfaces to compute the partition function of Yang-Mills theory (2dYM) on them. We find that the 
partition function depends on a combination of the two parameters that characterize a surface,
and therefore, distinguishes between these surfaces. 

Our analysis extends straightforwardly to 
the $q$-deformed generalization of Yang-Mills, and more generally to any two dimensional theory 
of topological character. In this article, we will only consider the case of two dimensional 
Yang-Mills theory in detail.  

\section{Yang-Mills theory on finite genus surfaces}
\label{sec:introfinite}
In this section we give a quick introduction to the Yang-Mills theory on finite genus surfaces with 
or without Wilson loops. We assume the gauge group $G$ to be compact and simply connected, and 
the Haar measure of $G$  to be normalized such that the volume of the group is
unity. Moreover, we denote the product $ag^{2}$, where $a$ is the area of the surface and $g$ is 
the coupling constant of Yang-Mills theory, simply as $a$ and refer to it as the area. With these 
assumptions the partition function for the disc of area $a$ is given as~\cite{Mig,Wit:first}
\begin{equation}
{Z(U;a)=\displaystyle\sum_{R}\dim R\: \exp(-aC_{2}(R))\chi_{_{R}}(U)}
\end{equation}
In the above, the sum is over all irreducible representations of the Lie group $G$, and $C_{2}(R)$ 
and $\chi(R)$ are, respectively, the second Casimir invariant and the character corresponding to the 
irreducible representation $R$. Lastly, $U$ is the holonomy along the boundary of the disc. The
partition function of any other surface is obtained by constructing the surface by gluing discs 
together and using the following integration identities~\cite{Ramg}
\begin{eqnarray}
{\displaystyle \int}_{G}dA\,\chi_{R_{1}}(BA)\,\chi_{R_{2}}(A^{-1}C) &=&
\delta_{R_{1},R_{2}}\,{\displaystyle \frac{\chi_{R_{1}}(BC)}{\dim R_{1}}}\,,\nonumber\\
{\displaystyle \int}_{G}dA\,\chi_{R}(ABA^{-1}C) & =& {\displaystyle 
\frac{\chi_{R}(B)\,\chi_{R}(C)}{\dim R}},\label{eq:group_id}
\end{eqnarray}
where integrations are with respect to the Haar measure of group $G$. 

Using the above formulae, the partition function for a surface of area $a$, genus $g$ and number 
of boundary components $p$ can be easily computed and is given as
\begin{equation}
Z^{\left(g,p\right)}(U_{1},\cdots,U_{p};a) = \displaystyle\sum_{R}(\dim R)^{2-2g-p}\; 
\exp\left(-aC_{2}�\right))\chi_R(U_{1}) \cdots\chi_R(U_{p}),
\label{MigdalForm} 
\end{equation}
where the holonomies $U_1,\cdots, U_p$ are associated with the $p$ boundaries. As is clear 
from this equation, by taking the infinite genus limit one is only left with the term corresponding to the 
one-dimensional representation. However, as we will see below, there is a class of infinite genus 
surfaces (namely Richards surfaces) for which the partition function has a nontrivial expression. 

More generally, we can insert Wilson loops on a surface. Consider, for example, the simplest 
situation in which a Wilson loop in representation $R_w$ is inserted along the nontrivial cycle 
of a cylinder, dividing it into two parts of areas $a_1$ and $a_2$ respectively. The corresponding 
partition function (with this loop insertion) is given as 
\begin{equation}
%\begin{split}
Z(U_{1},U_{2}; R_w ; a_1,a_2) = \displaystyle\sum_{R_1, R_2} N^{R_1}_{R_2 R_w} 
\exp\left(- a_1 C_{2}(R_1) - a_2 C_{2}(R_2)\right)
%\\ & \qquad\qquad\times\;
\,\chi_{R_1}(U_{1})\, \chi_{R_2}(U_{2}),
%\end{split}
\label{MigdalFormWilson} 
\end{equation}
where, the number $N^{R_1}_{R_2 R_w}$  is the multiplicity of the representation $R_1$ in the 
tensor product of $R_2$ and $R_w$. The partition functions of all finite genus surfaces with 
non-intersecting Wilson loops can be computed from Eqs.~\eqref{MigdalForm} and
\eqref{MigdalFormWilson} by using gluing techniques. 

%%%%%%%%%%%%%%%%%%%%%%%%%%%%%%%%%%%%
\section {Construction of Richards surfaces}
\label{sec:RichCons}
A surface is a connected orientable two dimensional topological manifold. 
A bordered surface $\Sigma$ is said to be of infinite genus if it does not have any bounded 
subset $S$ such that $\Sigma - S$ is of genus zero. A theorem in Ref.\cite{Brahana} classifies 
compact bordered surfaces: Two compact triangulable bordered surfaces are homeomorphic 
if and only if they both have  the same number of boundary curves, the same Euler characteristic, 
and are either both orientable or else both non-orientable. This was extended to the case of 
non-compact triangulable surfaces, including surfaces of infinite genus, by Ker\'{e}kj\'{a}rt\'{o}
and further generalized in Ref.\cite{Richards} (for surfaces without boundaries) and in 
Ref.\cite{PrMi} (for surfaces with boundaries).

Let us digress to define a few useful concepts. A subset of a surface $\Sigma$ is said to be 
{\em bounded} in $\Sigma$ if its closure in $\Sigma$ is compact. Let $P_{1}\supset P_{2}
\supset\cdots$ be a nested sequence of connected unbounded regions in $\Sigma$ such that 
the following holds: 
\begin{enumerate}
\item[(a)]
the boundary of $P_{n}$ in $\Sigma$ is compact for all $n$,
\item[(b)]
$P_{n}\cap A =\emptyset$ for $n$ sufficiently large for any bounded subset $A$ of $\Sigma$.
\end{enumerate}
Two sequences $P_{1}\supset P_{2}\supset\cdots$ and $Q_{1}\supset Q_{2}\supset\cdots$ are 
considered to be equivalent if for any $n$ there is an integer $N(n)$ such that $P_{n}\subset Q_{N}$ 
and vice versa. The equivalence class of a sequence $p=(P_{1}\supset P_{2}\supset\cdots)$ is  
called an {\em end} and will be denoted by $p^*$. In order to define a topology on the set of ends,
we start with a set $U$ in $\Sigma$, the boundary of which in $\Sigma$ is compact. Let us define 
open sets $U^{*}$ to be the set of all ends $p^{*}$, represented by some $p=(P_{1}\supset P_{2}
\supset\cdots)$, such that all except a finite number of $P_{n}\subset U$. An {\em ideal boundary} 
of a surface $\Sigma$ is the set $B(\Sigma)$ of its ends endowed with this topology.

Equivalently, let $\Sigma^*$ be a locally connected compactification of a non-compact surface 
$\Sigma$. The set $B(\Sigma) = \Sigma^* - \Sigma$ is the ideal boundary of $\Sigma$ if it is a 
totally disconnected set.

Two separable surfaces $\Sigma$ and $\Sigma'$ of the same genus and orientability class 
are homeomorphic if and only if their {\em ideal boundaries} are topologically 
equivalent\cite{Richards,PrMi}.

\bigskip

We would like to review the construction of an infinite genus orientable surface following
the ideas of Richards\cite{Richards}.
Let us start by recalling the construction of a Cantor set. We start with the closed line interval 
$[0,1]$, and repeatedly remove, say, the middle one-third from it. Therefore, after the first $n$ 
steps, we are left with $\left[0,\displaystyle\frac{1}{3^n}\right] \cup \left[\displaystyle\frac{2}{3^n},
\displaystyle\frac{1}{3^{n-1}}\right] \cup \cdots \cup \left[\displaystyle\frac{3^n-1}{3^n},1\right]$. 
The points that remain in the limit $n\to\infty$ constitute a Cantor set. However, more than just 
the end-points of the intervals that have been removed, remain in the limit. 
Indeed, even though the total length of the intervals that have been removed equals that of the 
original interval $[0,1]$, the cardinality of Cantor set is that of the closed set $[0,1]$. It is a totally 
disconnected space in which every point is an accumulation point. 

A Cantor set may also be constructed by removing every alternate $\ell$ intervals out of $2\ell+1$ 
segments for any positive integer $\ell$. However, Cantor sets associated with  different $\ell$ are 
isomorphic.
 
The iterative procedure for the construction of a Richards surface is related to a Cantor set. First, 
we start with a closed disc ${\cal D}_0$ containing\footnote{In Ref.\cite{Richards}, the points of the 
Cantor set are removed from the disc. For our purposes, from now on we will be working with a 
generalized notion of a surface that may include singular points.} $[0,1]$. In it are the disjoint 
discs ${\cal D}_{00}$ and ${\cal D}_{01}$ containing the intervals 
$\left[0,\frac{1}{3}\right]$ and $\left[\frac{2}{3},1\right]$, 
respectively. Now consider complement of (the union of) these smaller discs ${\cal D}_{00}$ and 
${\cal D}_{01}$ in ${\cal D}_{0}$. Attach $g$ handles to this part. That is, remove $2g$ disjoint 
discs and identify the boundaries of each pair preserving orientation. In the second step, similar 
operations are carried out in ${\cal D}_{00}$ and ${\cal D}_{01}$. After this process is repeated 
{\em ad infinitum}, it results in an infinite genus surface with $p=2$ branches and $g$ handles at 
each stage of branching. This is what we refer to as a {\em Richards surface} with parameters $p$ 
and $g$ (see Fig.~\ref{fig:RichardsTree}). The limiting intervals ${\cal D}_{000\cdots}$, 
${\cal D}_{001\cdots}$, ${\cal D}_{010\cdots}$, etc.\ are the boundary ends at infinity, where 
infinitely many handles converge, and hence are singular points of the surface. The boundary 
of the initial disc ${\cal D}_0$ is the only boundary of a Richards surface.  

By construction, the set of ends or ideal boundaries is homeomorphic to the Cantor set. If points 
of this Cantor set are {\em removed} from the surface (i.e., if the ends are punctured), then according 
to Richards' classification we get (up to homeomorphism) a unique  infinite genus surface which is 
independent of the choices of integers $g$ and $p$ made in the construction above. However, 
to our knowledge, it is not known whether different Richards surfaces (in our terminology) are 
homeomorphic or not. In the following we will use recursive nature of these surfaces to compute 
the partition function of Yang-Mills theory on these surfaces. We will find that Richards surfaces 
with different parameters have generically different partition functions.
%, and hence we can conclude 
%(at least at the level of rigour normally acceptable in Physics) that they are topologically inequivalent. 

%%%%%%%%%%%%%%%%%%%%%%%%%%%%%%%%%%%%%
\section{Yang-Mills Theory on Richards surfaces}\label{sec:YMonRichSur}
The self-similar structure of a Richards surface of infinite genus can be used to compute the 
partition function of Yang-Mills theory on it. 

%%%%%%%%%
\begin{figure}[tbp]
\centering
\begin{subfigure}{.5\textwidth}
  \centering
  \includegraphics[scale=0.43]{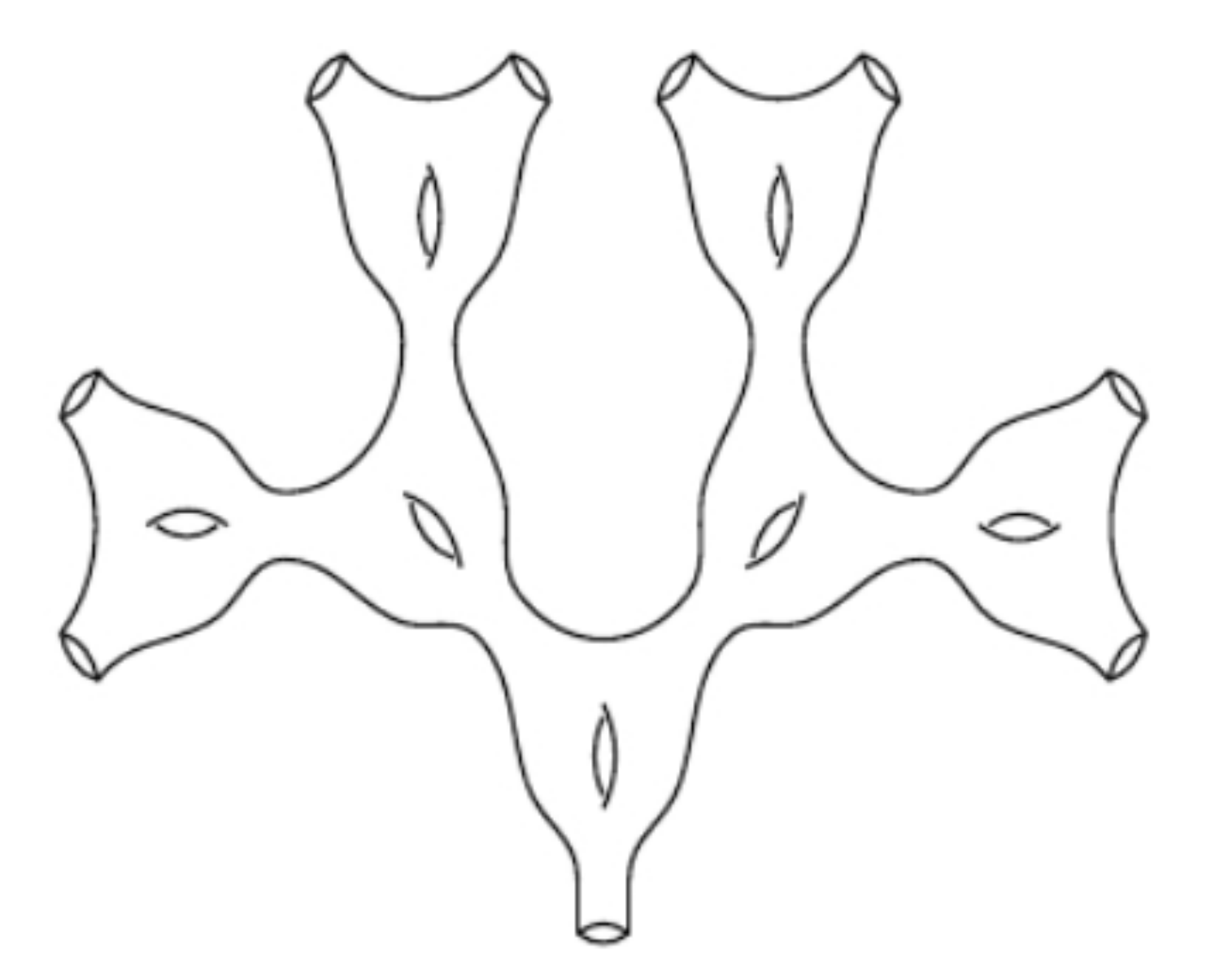}
  \caption{}
  \label{fig:RichardsTreeA}
\end{subfigure}%
\begin{subfigure}{.5\textwidth}
  \centering
  \includegraphics[scale=0.43]{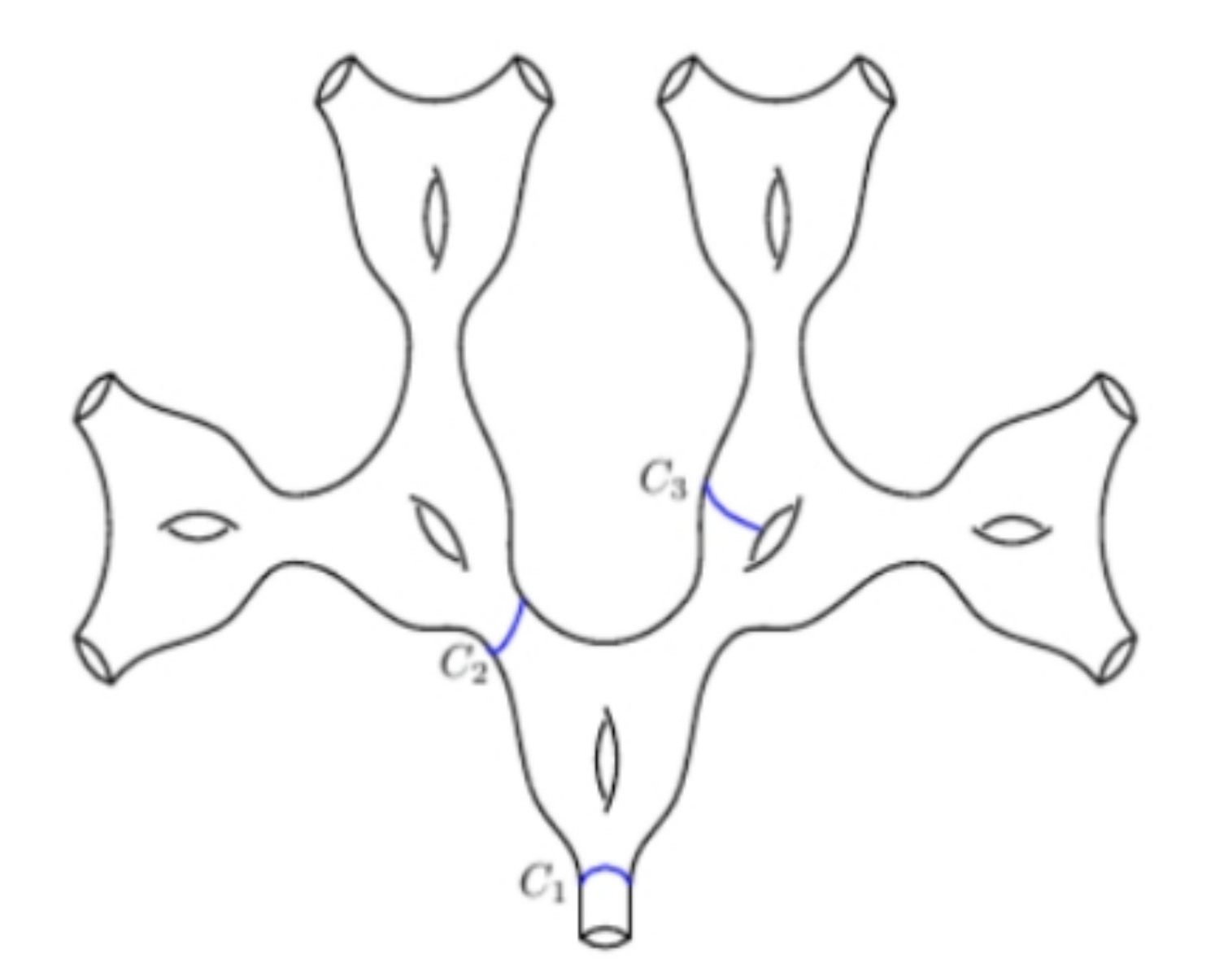}
  \caption{}
  \label{fig:RichardsTreeB}
\end{subfigure}
\caption{A Richards surface with parameters $p=2$ and $g=1$ (shown here upto level $n=3$) 
(a) without Wilson loops 
(b) with Wilson loops along the cycles $C_1$, $C_2$ and $C_3$.}
\label{fig:RichardsTree}
\end{figure}
%%%%%%%%%

%%%%%%%%%%%%%%%%%%%%%%%%%%%

First we consider a surface without insertion of additional Wilson loops. Let us associate a holonomy 
$U_0$ to the base boundary $\partial{\cal D}_0$, which is the only proper boundary of the surface. 
If we cut a Richards surface after $n$ branchings, we get a surface of (finite) genus 
$h=g\left(\displaystyle{\frac{p^n-1}{p-1}}\right)$ with $p^n$ copies of surfaces of infinite genus, 
each of which is isomorphic to the original Richards surface and are glued to the $p^n$ extreme boundaries. 
Therefore, we arrive at a recursion relation, which schematically reads as follows.
\begin{equation}
Z^{\left(g,p\right)}_\infty\left(U_0;a\right)\, = \, Z^{\left(g,p^n\right)}\left(U_0,U_1,\cdots, 
U_{p^n};a_1\right)\,  \displaystyle\prod_{i=1}^{p^n} Z^{\left(g,p\right)}_\infty
\left(U_i^\dagger;a-a_1\right).\label{RecRelRichards}
\end{equation}
In order to solve it, let us note that  $Z$, being invariant under $U \rightarrow hUh^{-1}$ is a 
class function. Therefore, we can expand it in terms of the basis $\chi_R$ of class functions as 
$Z^{\left(g,p\right)}_\infty\left(U_0;a\right)\, = \sum_{R} h(R,a) \chi_R (U_0)$, where $h(R,a)$ is a 
function that satisfies the recursion relation
\begin{equation}
h(R, pa+b) = (\mathrm{dim } R)^{1-2g-p} e^{-b C_2( R ) } h(R,a)^p.  \label{hRecursion}
\end{equation}
The unique solution to this is $h(R,a) = (\mathrm{dim } R)^{1 + \frac{2g}{p-1}}\, 
\exp\left(-a C_2( R ) \right)$. To see the uniqueness, consider taking $a=b=0$ in the recursion 
relation above. This determines, $h(R, 0) = (\mathrm{dim } R)^{1 + \frac{2g}{p-1}}$. Now taking 
$a=0$ and using $h(R,0)$ on the RHS of Eq.\refb{hRecursion}, we arrive at the solution of $h$. 
(We may also arrive at the same solution if we assume that it has the same functional form as 
Eq.~\eqref {MigdalForm}.) It is easy to see that the solution is independent of the level $n$ at 
which we choose to cut the Richards surface.

Therefore, the partition function on a surface with parameters $g$ (for the numbers of 
handles attached at each branch) and $p$ (for the number of branchings) is
\begin{equation}
Z^{\left(g,p\right)}_\infty\left(U_0;a\right)\, = \, \sum_R \left(\mathrm{dim}\, R\right)^{1 + \frac{2g}{p-1}}\,
e^{- a C_2(R)}\; \chi_R(U_0) . \label{pfYMRichards}
\end{equation}
(The above is not applicable to $p=1$, which corresponds to a surface known as the Loch Ness 
monster.)
Notice that, unlike in Eq.~\refb{MigdalForm}, the power of dim $R$ in the expression above, 
is a fraction $\alpha = 1 + \frac{2g}{p-1}$. However, it reduces to the known results in the special 
cases (i)  $g=0$, which is a disc with a holonomy $U_0$ at its boundary, and (ii) $p=0$, which is 
a surface of (finite) genus $g$ (with one boundary). 

We know that the power of dim $R$ in Eq.~\refb{MigdalForm} is the Euler characteristic of the surface. 
Since Richards surface is constructed by gluing (an infinite number of ) surfaces of finite characteristic,
its Euler characteristics is divergent. Nevertheless, the power of dim $R$ for the partition function of 
Richards surface in Eq.~\refb{pfYMRichards} may be thought of as the formal sum (disregarding
issues of convergence of the geometric series) of the Euler characteristics of its buildings blocks.

\bigskip

Now, consider a Richards surface in which we insert finitely many Wilson loops in certain 
representations $R_1,\cdots,R_n$ of the gauge group $G$  (see Fig.~\ref{fig:RichardsTreeB}). We 
can cut this surface at a finite level thus dividing it into (i) a surface of finite genus containing all the 
Wilson loops and (ii) a finite number of Richards surfaces without Wilson loops. Since the partition 
function of each part is known, the partition function of the total surface can be computed by the 
gluing method. Just as in the case of finite genus surfaces, its not difficult to verify that the final form 
of the partition function is independent of how the surface is cut into various sub-surfaces. 

\bigskip

Finally, the same method can also be applied to compute the partition function of the {\em 
quantum deformed} Yang-Mills theory on a Richards surface. Since the approach is exactly the 
same, we do not carry out the explicit calculations here. The result for the partition function in the 
$q$-deformed case can be obtained from Eq.~\refb{pfYMRichards} by obvious substitutions 
for the dimension and Casimir of the representations.

%%%%%%%%%%%%%%%%%%%%%%%%%%%%%%%%%%%%%
\section {Summary}
\label{sec:Concl}
In this note we computed the partition function of two dimensional Yang-Mills theory for a class 
of infinite genus surfaces. These surfaces are special in that they have a recursive structure, which 
could be exploited to solve for the partition function of the theory. Nevertheless, the fact that the 
partition function of a gauge theory on surfaces of infinite characteristics is a well behaved function, 
is somewhat of a surprise. We will analyze the analytic properties of this partition function as a 
function of the area elsewhere\cite{GIK}. It would be of interest to explore the structure of gauge 
theories, and more generally, topological theories, on these (or similar) recursive surfaces of 
infinite characteristics.

%%%%%%%%%%%%%%%%%%%%%%%%%%%%%%%%%%%%%%%%
%\bigskip

\acknowledgments
It is a pleasure to thank Debashis Ghoshal, Rajesh Gopakumar, Camillo Imbimbo, Ravi 
Prakash and Preeda Patcharamaneepakorn. This work was supported by a CSIR Research 
Fellowship.

%%%%%%%%%%%%%%%%%%%%%%%%%%%

\end{document}